\documentclass[%
 reprint,
superscriptaddress,
 amsmath,amssymb,
 aps,
 prl,
]{revtex4-2}

\usepackage{graphicx}% Include figure files
\usepackage{todonotes}
\usepackage{float}
\usepackage[version=4]{mhchem}
\usepackage{dcolumn}% Align table columns on decimal point
\usepackage{bm}% bold math
\usepackage{hyperref}% add hypertext capabilities
\hypersetup{
   breaklinks=true,   % splits links across lines
   colorlinks=true,   % displays links as colored text instead of blocks
}
\usepackage{siunitx}

\begin{document}
\preprint{APS/123-QED}
%\title{ High-Precision Mass Measurements of $^{52}$Ni and $^{51}$Co: Implications for Models of Mirror Energy Differences and the Isobaric Multiplet Mass Equation}
\title{ High-Precision Mass Measurements of $^{52}$Ni and $^{51}$Co Reveal Breakdown of the Isobaric Multiplet Mass Equation in the $fp$ Shell}
%T=5/2 mirror displacement energies and comparison to precision shell-model calculations}%

\author{F.~M.~Maier}
\email{franziska.m.maier@jyu.fi}
\affiliation{Facility for Rare Isotope Beams, East Lansing, Michigan, 48824, USA}
\author{G.~Bollen}
\affiliation{Facility for Rare Isotope Beams, East Lansing, Michigan, 48824, USA}
\affiliation{Department of Physics and Astronomy, Michigan State University, East Lansing, Michigan 48824, USA}
\author{B.~A.~Brown}
\affiliation{Facility for Rare Isotope Beams, East Lansing, Michigan, 48824, USA}
\affiliation{Department of Physics and Astronomy, Michigan State University, East Lansing, Michigan 48824, USA}
\author{S.~E.~Campbell}%
\affiliation{Facility for Rare Isotope Beams, East Lansing, Michigan, 48824, USA}
\affiliation{Department of Physics and Astronomy, Michigan State University, East Lansing, Michigan 48824, USA}
\author{X.~Chen}
\affiliation{Facility for Rare Isotope Beams, East Lansing, Michigan, 48824, USA}
\author{H.~Erington}
\affiliation{Facility for Rare Isotope Beams, East Lansing, Michigan, 48824, USA}
\affiliation{Department of Physics and Astronomy, Michigan State University, East Lansing, Michigan 48824, USA}
\author{N.~D.~Gamage}
\affiliation{Facility for Rare Isotope Beams, East Lansing, Michigan, 48824, USA}
\author{K.~Godbey}
\affiliation{Facility for Rare Isotope Beams, East Lansing, Michigan, 48824, USA}
\author{C.~M.~Ireland}
\affiliation{Facility for Rare Isotope Beams, East Lansing, Michigan, 48824, USA}
\affiliation{Department of Physics and Astronomy, Michigan State University, East Lansing, Michigan 48824, USA}
\author{C.~Izzo}%
\affiliation{Facility for Rare Isotope Beams, East Lansing, Michigan, 48824, USA}

\author{R.~Ringle}
\affiliation{Facility for Rare Isotope Beams, East Lansing, Michigan, 48824, USA}
\affiliation{Department of Physics and Astronomy, Michigan State University, East Lansing, Michigan 48824, USA}
\author{C.~S.~Sumithrarachchi}
\affiliation{Facility for Rare Isotope Beams, East Lansing, Michigan, 48824, USA}
\author{A.~C.~C.~Villari}
\affiliation{Facility for Rare Isotope Beams, East Lansing, Michigan, 48824, USA}
\date{\today}%

\begin{abstract}
\noindent
We performed high-precision mass measurements of the proton-rich nuclei $^{52}$Ni and $^{51}$Co with the LEBIT Penning trap at the Facility for Rare Isotope Beams (FRIB). For $^{52}$Ni, a mass excess of $-22474.8(2.2)$~keV was determined, which is consistent with a recent storage-ring measurement at the  Cooler-Storage Ring (CSRe) but has a factor 37 improved precision. For $^{51}$Co, we obtained a mass excess of $-27375.1(5.7)$~keV, agreeing with a recent CSRe result, while reducing the uncertainty by a factor~2. Combining our mass value for $^{52}$Ni with the known two-proton decay energy of $^{54}$Zn, we determined the mass excess of $^{54}$Zn to be $-6463(42)$~keV. These new mass values reveal a substantial breakdown of the isobaric mass multiplet equation for $A=52$ and $A=54$, and provide stringent benchmarks for isospin-symmetry-breaking effects in the proton-rich $fp$-shell, favoring theoretical descriptions that omit the Coulomb-exchange term.  
\end{abstract}

\maketitle

\section{I. Introduction}
The isobaric multiplet mass equation (IMME) provides a powerful framework for predicting excitation energies and ground-state masses of exotic nuclei that are challenging to measure experimentally. It plays an important role in nuclear physics and astrophysics, providing mass predictions that support stellar evolution modeling~\cite{PhysRevC.79.045808, PhysRevC.65.045802, PhysRevC.95.055806} and aides in determining the position of the proton dripline, which is crucial for understanding the boundaries of nuclear stability~\cite{sym16060745}. It furthermore provides information on the vector and tensor components of isospin non-conserving forces and is used to test the conserved vector current hypothesis of the electroweak interaction~\cite{PhysRevLett.94.092502, PhysRevC.91.025501}. The IMME relates the binding energies (BE) of the $2T+1$ members of an isobaric multiplet through
\begin{equation}
    \mathrm{BE}(\alpha,T,T_z)=a(\alpha,T)+b(\alpha,T)T_z+c(\alpha,T)T_z^2, \label{eq: IMME}
\end{equation}
where $a$, $b$ and $c$ are fitting coefficients and $\alpha$ specifies the spin-parity of the nuclear state~\cite{PhysRev.116.465}. The $2T+1$ members of an isobaric multiplet, commonly known as isobaric analog states, share the same isospin $T$, mass number $A$ and spin-parity $J^\pi$, while differing in their isospin projection $T_z$. 
The $b$-coefficient can also be calculated from the differences in binding energy of mirror nuclei,
\begin{equation}
    b= \frac{\mathrm{BE}(T, T_z=T) - \mathrm{BE}(T, T_z=-T)}{2T} \label{eq: b coeff}
\end{equation}
and is sensitive to the isovector component of the nuclear Hamiltonian. Consequently, the $b$-coefficient provides a stringent benchmark for nuclear energy-density functionals and for their treatment of Coulomb exchange and other isospin-symmetry-breaking effects. Higher order terms of the IMME, such as the cubic $dT_z^3$ and quartic $e T_z^4$ terms, are for $T>1$ found to be small~\cite{MACCORMICK201461}. 

High-precision mass measurements of proton-rich nuclei provide critical tests of the validity of the IMME as well as of theoretical descriptions of isospin-symmetry-breaking effects, see e.g. recent work~\cite{14s5-17gj, PhysRevC.107.014302, physics5020026, zy1j-mynp}. In the $fp$-shell, where Coulomb interactions become increasingly significant, deviations between predicted and measured IMME coefficients may reveal deficiencies in the underlying treatment of the Coulomb interaction~\cite{PhysRevLett.109.102501}. 

In this work, the masses of $^{52}$Ni and $^{51}$Co were measured at the Low Energy Beam and Ion Trap (LEBIT)~\cite{LEBIT} at the Facility for Rare Isotope Beams (FRIB). The new measurements improve the precision of the previously known mass values~\cite{Zhang2023, PhysRevC.102.054311} by a factor of 37 and 2, respectively. Our results extend tests of the IMME further into the proton-rich $fp$-shell region, where discrepancies between calculated and experimental $b$-coefficients expose limitations in the treatment of Coulomb and other isospin-symmetry-breaking effects. For $A=52$, $T=2$ and $A=54$, $T=3$, the cubic correction terms of the IMME, $d T_z^3$, differ by 7$\sigma$ and 10$\sigma$ from 0 when taking our new mass values into account. These represent the largest known deviations from the quadratic form of the IMME alongside the $A=8, 9, 22, 32, 35$ and $44$ multiplets.

\section{II. Experimental Method and Analysis}
A radioactive beam containing $^{52}$Ni and $^{51}$Co was produced at FRIB via projectile fragmentation of a $^{58}$Ni primary beam. The primary beam was accelerated to an energy of 250 MeV/u using FRIB's superconducting linear accelerator~\cite{doi:10.1142/S0217732322300063} and subsequently impinged on a 5~mm thick $^\mathrm{nat}$C production target, generating a broad distribution of reaction products. The ions of interest were separated from the unwanted fragments and the remaining primary beam by the Advanced Rare Isotope Separator (ARIS)~\cite{PORTILLO2023151} and transported to the Advanced Cryogenic Gas Stopper (ACGS)~\cite{Lund-ACGS}. In the final energy-dispersive beamline section leading to the ACGS, the beam momentum distribution of the purified beam was compressed using a 1514~$\mu$m thick Al wedge, and a 1395~$\mu$m thick Al degrader with an inclination angle of 29~degrees reduced the momentum. 

In the ACGS, the beam was stopped via collisions with helium buffer gas and guided towards the extraction orifice using radio-frequency (RF) ion-carpet surfing~\cite{BOLLEN-IonSurfing}. Following extraction, the ions passed through an RF quadrupole that served as both a beam cooler and differential pumping barrier. The ions were subsequently accelerated to a beam energy of 30 keV and mass-separated using a dipole magnet with a mass resolving power of $\approx 1500$. The resulting continuous beam was transmitted to LEBIT. 
At LEBIT, the ions were then delivered to a linear Paul-trap cooler-buncher~\cite{CoolerBuncher}, utilizing helium buffer gas to accumulate, cool, and bunch the beam. After a cooling period of $10$~ms, the ions were extracted from the cooler-buncher as low-emittance ion bunches and guided into LEBIT's 9.4~T hyperbolic Penning trap~\cite{PenningTrap}.

In the Penning trap, ions were confined in all three spatial dimensions by a superposition of a magnetic field~$B$ and an electrostatic quadrupole field. The ion motion is characterized by three eigenfrequencies: the axial frequency $\nu_{z}$ and two radial frequencies, the magnetron frequency $\nu_{-}$ and the reduced cyclotron frequency $\nu_{+}$. Under ideal conditions, the cyclotron frequency $\nu_{c}$ is given by the sum of the two radial frequencies, $\nu_{c}=\nu_{+} + \nu_{-}$~\cite{Gabrielse-Sideband} and is related to the ion's mass-to-charge ratio $m/q$ through 
\begin{equation}\label{eq:nuc}
  \nu_{c} = \frac{q}{m} \frac{B}{2\pi}.  
\end{equation}
To minimize the impact of magnetic-field drifts and systematic uncertainties, measurements of $\nu_{c}$ of the ions of interest were interleaved with measurements of a well-known reference species, $\nu_{c, \mathrm{ref}}$, thereby eliminating the dependence on $B$. The cyclotron-frequency ratio $R$ is given by
\begin{equation}\label{eq:mass_from_ratio}
   R = \frac{ \nu_{c} }{ \nu_{c,\textrm{ref}} } = \frac{ q \cdot m_\textrm{ref} }{ q_\textrm{ref}\cdot m },
\end{equation}
where $q$ and $q_\textrm{ref}$ are the charges and $m$ and $m_\textrm{ref}$ are the masses of the ion of interest and the reference ion, respectively. 
The mass of $^{51}$Co$^{2+}$ ions was measured relative to the reference ions N$_2^+$, while the mass of $^{52}$Ni$^{2+}$ was determined using C$_2$H$_2^+$ as the reference species. The atomic mass values of $^{51}$Co and $^{52}$Ni were then extracted from the measured frequency ratios $R$ accounting for the different charge states and respective electron masses. Binding energies of the electrons were neglected as their contributions are significantly smaller compared to the statistical uncertainty of $R$.

\begin{figure}[t]
\includegraphics[width=\columnwidth]{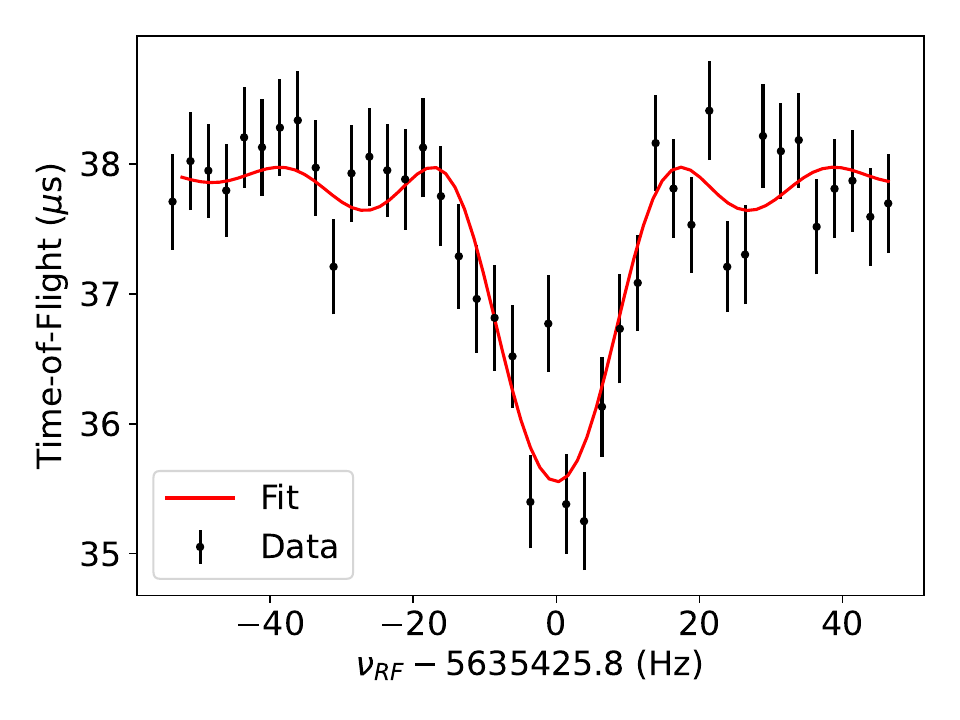}
\caption{A ToF-ICR spectrum of \textsuperscript{51}Co\textsuperscript{2+} measured with a quadrupole excitation time of 50~ms. The data of all performed measurements (measurement numbers 1-3 in Fig.~\ref{fig:Ratios}(a)) is combined. The spectrum is formed by 6614 detected ions. The red curve represents a $\chi^{2}$-minimization fit to the experimental data points shown in black, following the method  described in \cite{KONIG199595}. The cyclotron frequency $\nu_{c}$ is determined from $\nu_{RF}$ at the minimum time-of-flight in the fitted ToF-ICR spectrum. }
\label{fig:51Co_ToFICR_summed}
\end{figure}

\begin{figure}[t]
\includegraphics[width=0.8\columnwidth]{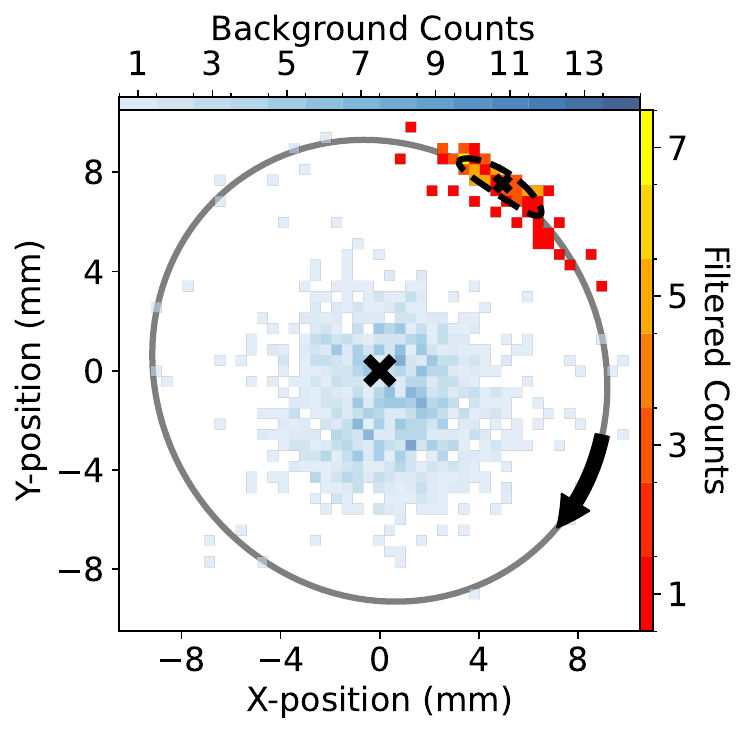}
\caption{The combined spatial distribution of the final two PI-ICR measurements of \textsuperscript{52}Ni\textsuperscript{2+} (measurements 3 and 4 in Fig.~\ref{fig:Ratios}(bottom)). The spot containing 71 \textsuperscript{52}Ni\textsuperscript{2+} ions is shown alongside unidentified background (likely C$_2$H$_2^+$).  A $\nu_+$ accumulation time of 31~ms was used.  The black arrow indicates the direction of the motion along the gray elliptical path, which center is marked with a large black X. The filtered \textsuperscript{52}Ni\textsuperscript{2+}data is fitted to a two-dimensional Gaussian distribution in polar coordinates, which fitted center and 1$\sigma$ uncertainty band are marked by a small black x and a dashed black ellipse, respectively. }
\label{fig:52Ni_PIICR_summed}
\end{figure}

For the mass measurement of $^{51}$Co$^{2+}$, the time-of-flight ion-cyclotron-resonance (ToF-ICR) technique~\cite{ToF1,BECKER199053,KONIG199595} was employed. The relatively high production rate and high beam purity of $^{51}$Co made the established ToF-ICR technique well suited for this measurement, without requiring the additional offline setup time of the phase-imaging ion-cyclotron-resonance (PI-ICR) method. After extraction from the Paul-trap cooler-buncher, the ions were displaced from the trap axis using a Lorentz steerer~\cite{LorentzSteerer}, thereby establishing an initial radial offset of the ions with respect to the trap center. Upon capture in the Penning trap, this offset induced a magnetron motion. Subsequently, a quadrupole RF excitation pulse with a frequency close to the expected cyclotron frequency of the ion of interest was applied, which converted the slow magnetron motion into a fast reduced cyclotron motion.  An excitation time of 500~ms was used for the stable reference N$_2^+$ ions. Owing to the short half-life of $^{51}$Co ($T_{1/2} = 68.8$ ms), the excitation time for $^{51}$Co$^{2+}$ was reduced to 50~ms.  Following the excitation, the ions were ejected from the Penning trap and their time-of-flight to a microchannel-plate (MCP) detector was recorded. The frequency of the quadrupole pulse $\nu_{\mathrm{RF}}$ was scanned, resulting in varying degrees of conversion between the magnetron and cyclotron motions. As $\nu_{\mathrm{RF}}$ approached the cyclotron frequency $\nu_c$, the radial kinetic energy of the ions increased, leading to shorter flight times, see Fig.~\ref{fig:51Co_ToFICR_summed}. The minimum in the time-of-flight spectrum occurred when $\nu_{\mathrm{RF}}$ equals $\nu_c$, allowing a precise determination of $\nu_c$.  Three cyclotron frequency measurements of $^{51}$Co$^{2+}$ were interleaved with measurements of the stable reference N$_2^+$ within a total measurement time of 1~hour 40~minutes. The rate of detected ions was capped at around 2 ions per second to minimize Coulomb interaction of the ions while stored in the Penning trap~\cite{ToF1}. 

For $^{52}$Ni$^{2+}$, the phase-imaging ion-cyclotron-resonance (PI-ICR) technique~\cite{PhysRevLett.110.082501, Eliseev2014} was used due to the substantially lower yield of $^{52}$Ni$^{2+}$ compared to $^{51}$Co$^{2+}$, following the procedure described in~\cite{vck7-1c4t, PhysRevLett.132.152501}. In this technique, the excited ion evolves for an accumulation time $t_\mathrm{acc}$ in the trap, before it is extracted onto a position-sensitive MCP detector. An example of the spatial ion distribution recorded on the MCP detector for a $\nu_+$ measurement of $^{52}$Ni$^{2+}$ is shown in Fig.~\ref{fig:52Ni_PIICR_summed} for $t_\mathrm{acc}=31$~ms. The radial frequencies $\nu_-$ and $\nu_+$ were measured independently and are given by
\begin{equation}
    \nu_\pm = \frac{\Delta \phi_{\mathrm{acc}_\pm} + 2\pi n_\pm}{2 \pi t_{\mathrm{acc}_\pm}},
\end{equation} where $\Delta \phi_{\mathrm{acc}_\pm}$ is the net phase advance during the time $t_{\mathrm{acc}\pm}$. The number of completed revolutions inside the trap $n_\pm$ is known when the frequency uncertainty is smaller than $1/t_\mathrm{acc}$. Since the $\nu_-$ motion only very weakly depends on the mass, the $\nu_-$ of $^{52}$Ni$^{2+}$ was linearly interpolated from that of the isobaric reference C$_2$H$_2^+$ enabling us to obtain higher statistics and precision of the $\nu_+$ measurement of $^{52}$Ni$^{2+}$, that only has a half life of 41.4~ms~\cite{Kondev_2021}. A phase accumulation time $t_{\mathrm{acc}_\pm}$ of 50~ms was chosen for the reference measurements, whereas for the ions of interest it was 5~ms for the first $\nu_+$ measurement, 13.5~ms for the second measurement and 31~ms for the last two measurements. The four $\nu_+$ measurements of $^{52}$Ni$^{2+}$ lasted 15, 4.5, 27.5 and 26.5~minutes each with 54, 19, 41 and 30 recorded $^{52}$Ni$^{2+}$ ions, respectively. 
Given the short 41.4~ms half-life of $^{52}$Ni and the 3-hour constraint for the measurement campaign of $^{52}$Ni, measurements with a higher accumulation time were not attempted.

\begin{figure}[t]
\includegraphics[width=\columnwidth]{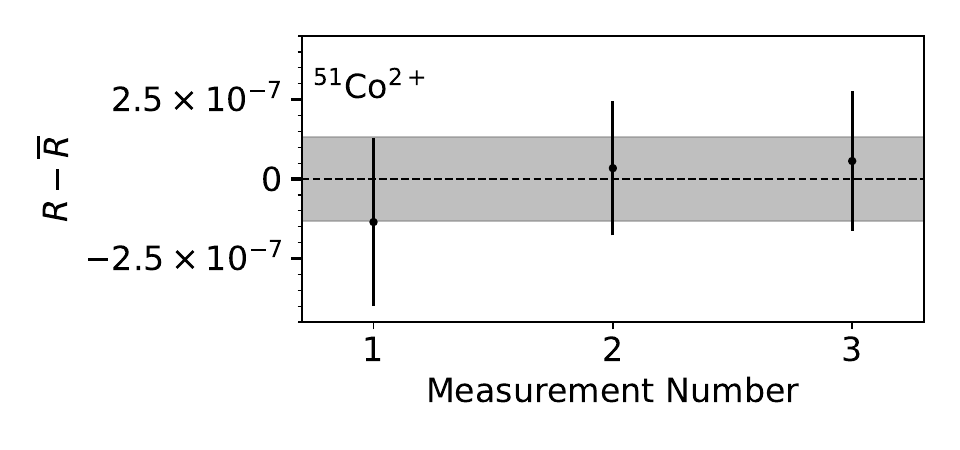}
\includegraphics[width=\columnwidth]{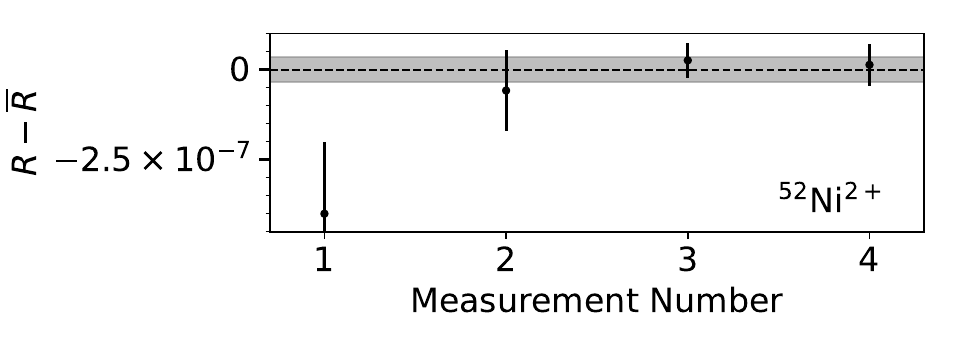}
\caption{Cyclotron frequency ratios $R$ relative to the weighted average value $\bar{R} = 1.09891568(14)$ for $^{51}$Co$^{2+}$ (top panel) and $\bar{R} = 1.001066436(35)$ for $^{52}$Ni$^{2+}$ (bottom panel). The gray shaded band indicates the $\pm 1 \sigma$ uncertainty in $\bar{R}$.}
\label{fig:Ratios}
\end{figure}

\section{III. Results }
The measured frequency ratios $R$ of $^{51}$Co$^{2+}$ and $^{52}$Ni$^{2+}$ relative to the weighted average cyclotron frequency ratios $\bar{R}=1.09891568(14)$ and $\bar{R}=1.001066436(35)$ are shown in Fig.~\ref{fig:Ratios}. 
The mass excess values of $^{51}$Co and $^{52}$Ni are -27375.1(5.7)~keV and -22474.8(2.2)~keV. Systematic errors as discussed in detail in previous work~\cite{MassOffsetError, MagneticFieldShift, vck7-1c4t, PhysRevLett.132.152501} are negligible compared to the statistical error. Only measurement cycles with less than 5 ions/bunch were used in the analysis to avoid any space-charge effects. No difference in the result was obtained when limiting the number of ions to $<2$ or $<7$ ions per bunch. For the PI-ICR measurements, we furthermore performed an averaging over the expected magnetron period of the ions prior to their excitation to mitigate initial residual magnetron motion that might originate when the ions are not injected perfectly onto the central axis of the Penning trap. The PI-ICR spots were fitted with a two-dimensional Gaussian distribution in polar coordinates (see Fig.~\ref{fig:52Ni_PIICR_summed}) as well as with a polar Skew-t distribution~\cite{BRANCO200199} to account for a slight tail in clockwise direction. Such tails are observed when the trap tuning is not perfect and may introduce mass-dependent shifts when the ions of interest and the reference ions have different $A/q$ values. In the present measurement, however, the $A/q$ values of $^{52}$Ni$^{2+}$ and the reference C$_2$H$_2^+$ were identical, making any such shift negligible. Consistent with this expectation, the frequency ratios extracted from the Gaussian and Skew-t fits agreed within their statistical uncertainties.

The measured cyclotron frequency ratios $R$ were compared with the expected $R$ for potential (molecular) isobaric contaminants. No candidate isobars were found, indicating that the detected ions were $^{51}$Co$^{2+}$ and $^{52}$Ni$^{2+}$. 

Our measured mass excess value of $^{51}$Co, $-27375.1(5.7)$~keV, represents a factor 2 improvement in precision compared to the most recent measurement performed at the Cooler-Storage Ring (CSRe) in Lanzhou~\cite{Zhang2023}. It is also in line with an earlier mass measurement from CSRe~\cite{SHUAI2014327, Zhou2021}, while being a factor 7 more precise, see Tab.~\ref{tab: MassExcessTable}. The mass excess of $^{52}$Ni, $-22474.8(2.2)$~keV, agrees with a recent CSRe measurement~\cite{PhysRevC.102.054311}, too, but improves the precision by a factor~37. 

\begin{table*}[t] 
\centering
\caption{Mass excess for $^{51}$Co, $^{52}$Ni and $^{54}$Zn, compared with the recent values by CSRe. The mass deviation $\Delta$ME is calculated by the difference of the mass excess values in this work and CSRe. The symbol \# stands for an extrapolated value in AME2020~\cite{AME2020}.}
\vspace{\baselineskip}
\renewcommand{\arraystretch}{1.25}
\setlength{\tabcolsep}{12pt}
\begin{tabular}{l c c c c}
\hline
& $\textrm{ME}_{\textrm{This work}}$ (keV) & $\textrm{ME}_{\textrm{AME2020}}$ & $\textrm{ME}_{\textrm{CSRe}}$ (keV) & $\Delta$ME (keV)  \\ \hline
$^{51}$Co   & $-27375.1(5.7)$ & $-27340(50)$& $-27342(48)$~\cite{SHUAI2014327}   & $33(49)$  \\ 
            &                 &   &  $-27332(41)$~\cite{Zhou2021}  & $43(42)$  \\ 
             &                 &   &  $ -27386(12)$~\cite{Zhang2023}  & $11(14)$  \\ 
$^{52}$Ni  & $ -22474.8(2.2)$ & $-22560(80)$  &   $-22560(83)$~\cite{PhysRevC.102.054311} & $85(84)$   \\ 
$^{54}$Zn  &$-6463(42) $      & $-5700(220)$\#  & $-6504(85)$~\cite{Li_2022}    & $41(95)$   \\ 
\hline
\end{tabular}
\label{tab: MassExcessTable}
\end{table*}

Based on our mass measurement of $^{52}$Ni and the two-proton decay energy $Q_{2p}$ of $^{54}$Zn, we can determine the mass excess of $^{54}$Zn, 
\begin{equation}
    \mathrm{ME}(^{54}\mathrm{Zn}) = \mathrm{ME}(^{52}\mathrm{Ni}) + 2\mathrm{ME}(^{1}\mathrm{H}) + Q_{2p} \label{eq:54Zn}.
\end{equation}
The two proton Q value, $Q_{2p}$, of $^{54}$Zn was measured at GANIL in 2005 with 1480(20)~keV~\cite{PhysRevLett.94.232501} and in 2011 with 1280(210) keV~\cite{PhysRevLett.107.102502}. These values were, however, replaced in the Atomic Mass Evaluation AME2020~\cite{AME2020} by an extrapolated value of 2280(200)~keV since they were not in agreement with expected trends of the mass surface. A measurement performed in 2026 at RIKEN led to $Q_{2p}=1363(25)$~keV~\cite{p8lf-hc4y}, showing a substantially better agreement with the two previous measurements than the extrapolated value. In this work, we adopt the two-proton decay energy of $^{54}$Zn with 1433(41)~keV, representing the weighted average of the two measurements at GANIL~\cite{PhysRevLett.94.232501, PhysRevLett.107.102502} and the measurement at RIKEN~\cite{p8lf-hc4y}. The uncertainty is inflated by their Birge ratio~\cite{PhysRev.40.207} of 2.6. Employing Eq.~\ref{eq:54Zn}, the mass excess of $^{54}$Zn follows as $-6463(42)$~keV. It differs by $763(224)$~keV from the extrapolated value in AME2020. A direct mass measurement of $^{54}$Zn would resolve this discrepancy. Due to the short half life of $1.59^{+60}_{-35}$~ms and the low production cross sections, such a mass measurement remains experimentally challenging. In the near future, a direct measurement will at FRIB likely only be feasible using the time-of-flight-magnetic-rigidity technique~\cite{MATOS2012171, MEISEL2013145}.

\section{IV. Discussion}

\begin{figure*}[t]
\includegraphics[width=0.8\textwidth]{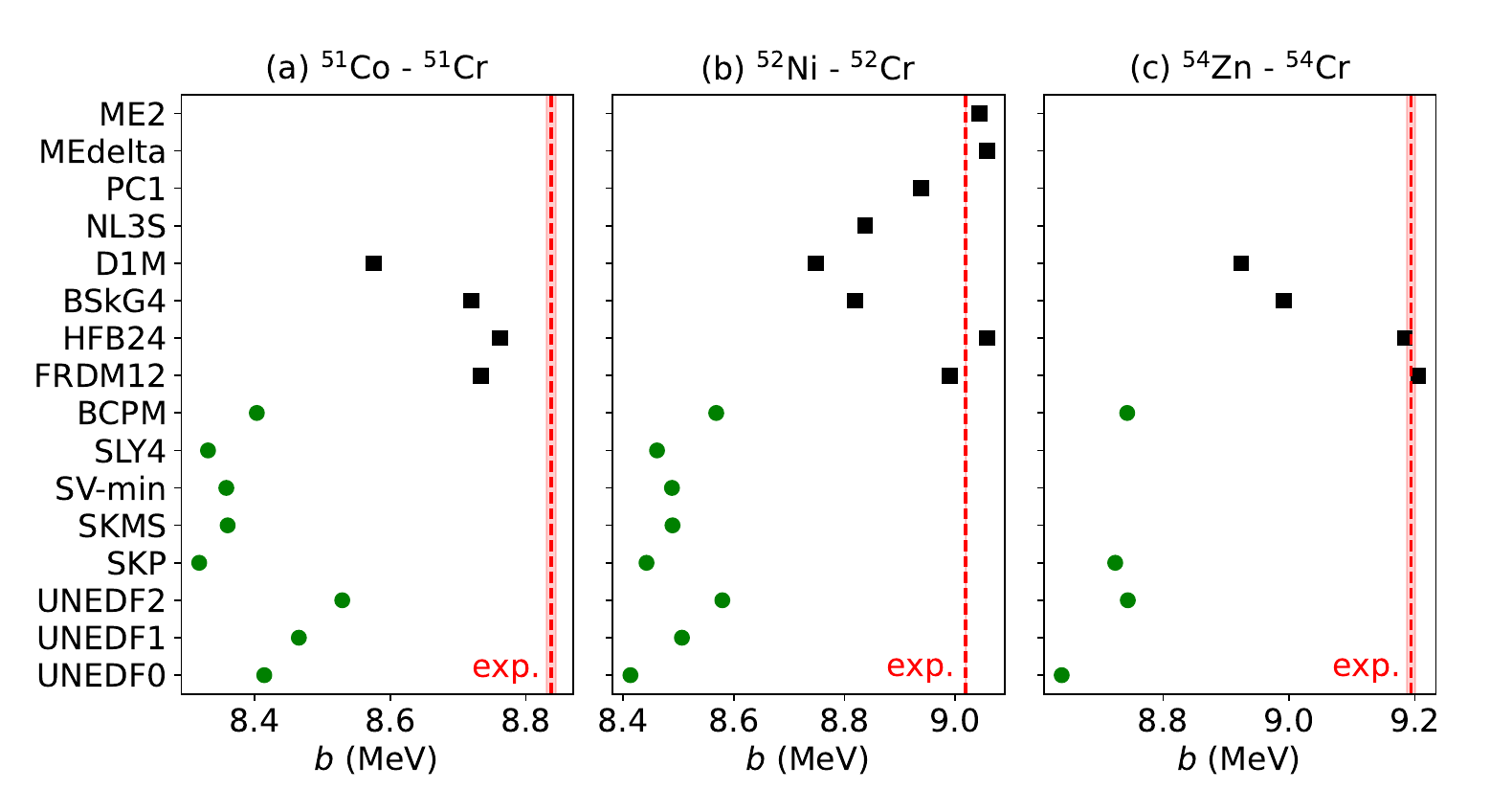}
\caption{Comparison of the $b$ coefficients of various theoretical models~\cite{dd-me2,dd-medelta,dd-pc1,nl3s,d1m,BSKG4,HFB24,HFB24dataset,FRDM2012,bcpm,SLY4,SVMIN,SKMstar,SKP,UNEDF2,UNEDF1,UNEDF0} with the experimental values for (a) $^{51}$Co - $^{51}$Cr, (b) $^{52}$Ni - $^{52}$Cr and (c) $^{54}$Zn - $^{54}$Cr. The experimental data is taken from this work and for the Cr isotopes from NUBASE2020~\cite{Kondev_2021}. For $A=51$, some models do not include calculations for odd-even nuclei. For $A=54$, some models do not include calculations of $^{54}$Zn. Black squares show model predictions without the Coulomb-exchange term, while green points represent predictions that include it.}
\label{fig:bvstheory}
\end{figure*}
Comparing the experimental and predicted $b$ and $d$ coefficients of the IMME, provides insights into Coulomb and other isospin-symmetry-breaking effects.
As a first step, we extract the $b$-coefficients for the mirror pairs $^{51}$Co - $^{51}$Cr, $^{52}$Ni - $^{52}$Cr and $^{54}$Zn - $^{54}$Cr using our new experimental mass values together with the mass values from the NUBASE2020 database~\cite{Kondev_2021} employing Eq.~\ref{eq: b coeff}. We then compare them with predictions from several global models in Fig.~\ref{fig:bvstheory}. Since the $b$ coefficients are primarily governed by Coulomb effects and other isospin-symmetry-breaking effects, they provide a sensitive probe of how these terms are treated in modern nuclear models.
For a closed-shell configuration, the Coulomb energy contains
a direct and an exchange term. Both terms can be calculated
exactly~\cite{PhysRevC.84.014310}. The exchange term reduces the total Coulomb energy by about 5\%.
The exact exchange term in Energy Density Functional (EDF) calculations is complex~\cite{PhysRevC.84.014310}, and
for practical purposes one usually uses the
Slater approximation~\cite{PhysRev.81.385}.
The ratio of the exact exchange calculation
over the Slater approximation goes from 1.08 for $^{16}$O
down to 1.02 for $^{208}$Pb \cite{PhysRevC.84.014310}.
The Coulomb energy can be isolated through the
difference in the binding energies of mirror nuclei, called displacement energies. These would be zero if the Coulomb interaction is turned off
and the strong interaction is charge independent, meaning that the proton-proton interaction is the same as the
neutron-neutron interaction.
The Slater approximation with charge independence
is used for all of the Skyrme-type \cite{PhysRevC.5.626} EDF functionals
shown with green dots in Fig~\ref{fig:bvstheory}.
The displacement energies for these are all lower than in the experiment
by about 5\%.

During the determination of the SKX EDF functional parameters \cite{PhysRevC.58.220},
it was observed that the calculated displacement energy
for the mirror pair $^{48}$Ni - $^{48}$Ca was in better agreement with experiment when the
exchange term is removed, e.g. just the direct Coulomb
energy term is used. All displacement energies
were analyzed in Ref.~\cite{BROWN200049} where it was found that
the EDF calculated values were in much better agreement
with experiment without the Coulomb-exchange term.
Comparable results can also be obtained if a charge-symmetry breaking
term is added to the Skyrme functional,
but the needed value is 3-4 times larger than
that expected from nucleon-nucleon interactions \cite{PhysRevC.58.220}.

All nuclear wavefunctions
contain correlations beyond the single Slater
determinant (closed-shell) configurations
used for the EDF basis~\cite{PhysRevC.58.220}.
These correlations could give corrections
to the Coulomb energy
that effectively cancels the Coulomb-exchange term \cite{BULGAC1996103, BULGAC19991, Bulgac1999}.
At the phenomenological level, this term is
added to a Kohn-Sham EDF functional \cite{PhysRev.140.A1133}.
For the HFB24 and BSkG4 Skyrme-type EDF functionals,
the Coulomb-exchange term is not included \cite{HFB24}.
Also, the Coulomb-exchange term is not included in
any of the Covariant mass models \cite{PhysRevC.93.054310}. These models, shown with black squares in Fig~\ref{fig:bvstheory}, significantly better reproduce the experimental data. The finite-range droplet model (FRDM12)~\cite{moller2016nuclear} also provides reasonably good agreement.

\begin{figure*}[t]
\includegraphics[width=\textwidth]{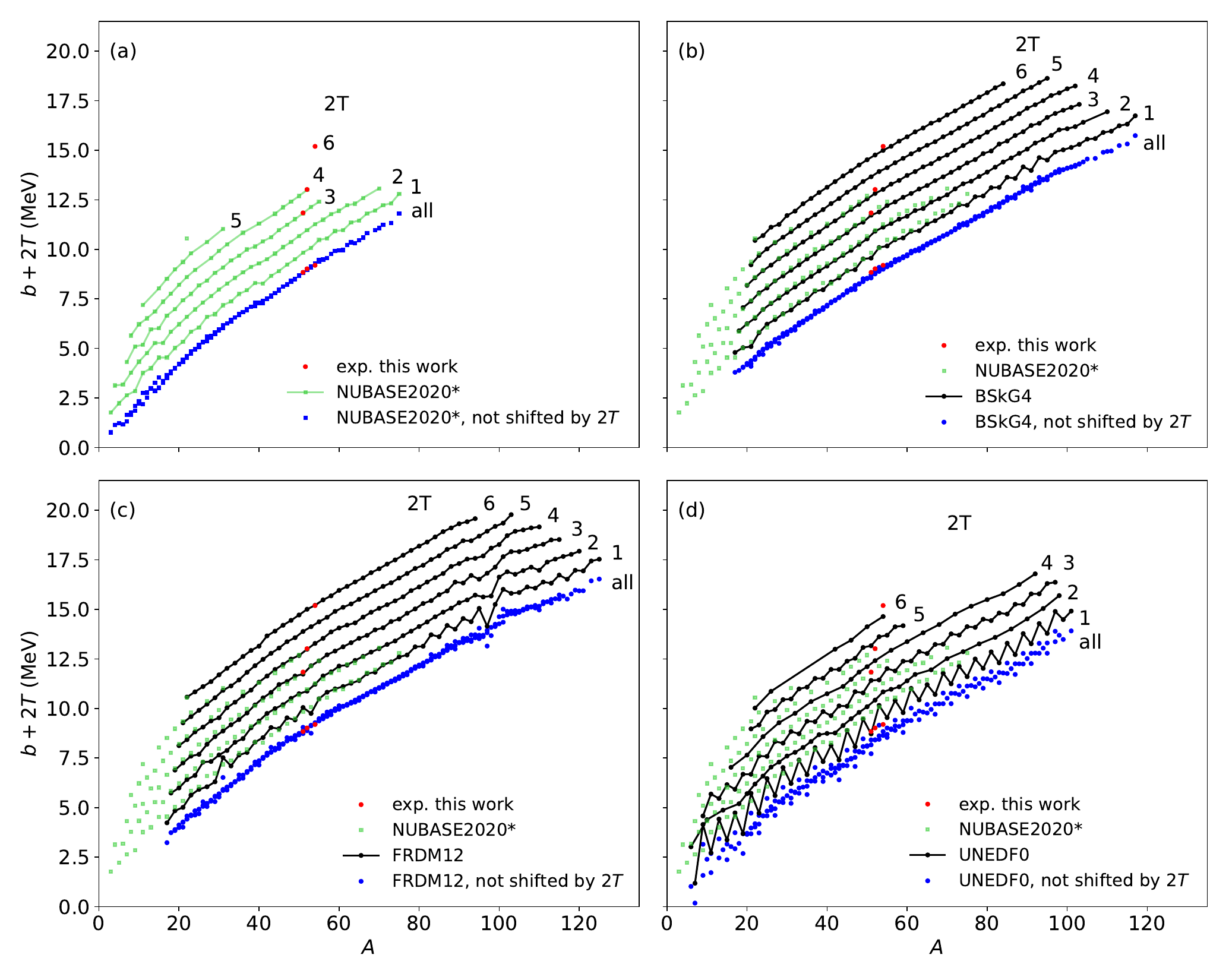}
\caption{The $b$-coefficients, shifted by $2T$, as a function of mass number across the nuclear chart.  Our results (red points) are compared with the data in NUBASE2020 including recent mass measurements as reported in Refs.~\cite{PhysRevLett.133.222501, 14s5-17gj, PhysRevC.110.L031301, ffwt-n7yc} in panel (a),  with the predictions of the BSKG4 model~\cite{BSKG4} in panel (b), with FRDM12~\cite{moller2016nuclear} in panel (c) and with UNEDF0~\cite{UNEDF0} in panel (d). The experimental data prior to this work is shown in green in all four panels.}
\label{fig:bvsA_fullchart}
\end{figure*}

\begin{figure}[t]
\includegraphics[width=\columnwidth]{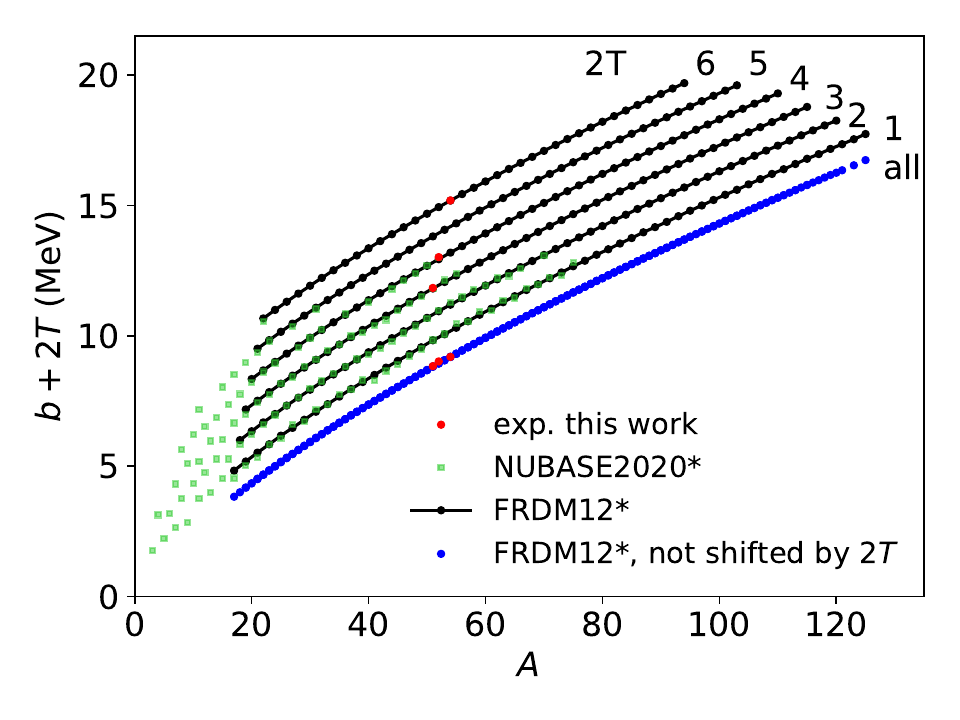}
\caption{The $b$-coefficients, shifted by $2T$, as a function of mass number across the nuclear chart.  Our results (red points) are compared with the predictions of the FRDM12* model~\cite{FRDM2012}, that neglects the $E_{\mathrm{mic}}$ term present in the FRDM12 model. The experimental data prior to this work shown in green is taken from NUBASE2020 including recent mass measurements as reported in Refs.~\cite{PhysRevLett.133.222501, 14s5-17gj, PhysRevC.110.L031301, ffwt-n7yc}. }
\label{fig:FRDMstar}
\end{figure}

\begin{figure}[t]
\includegraphics[width=\columnwidth]{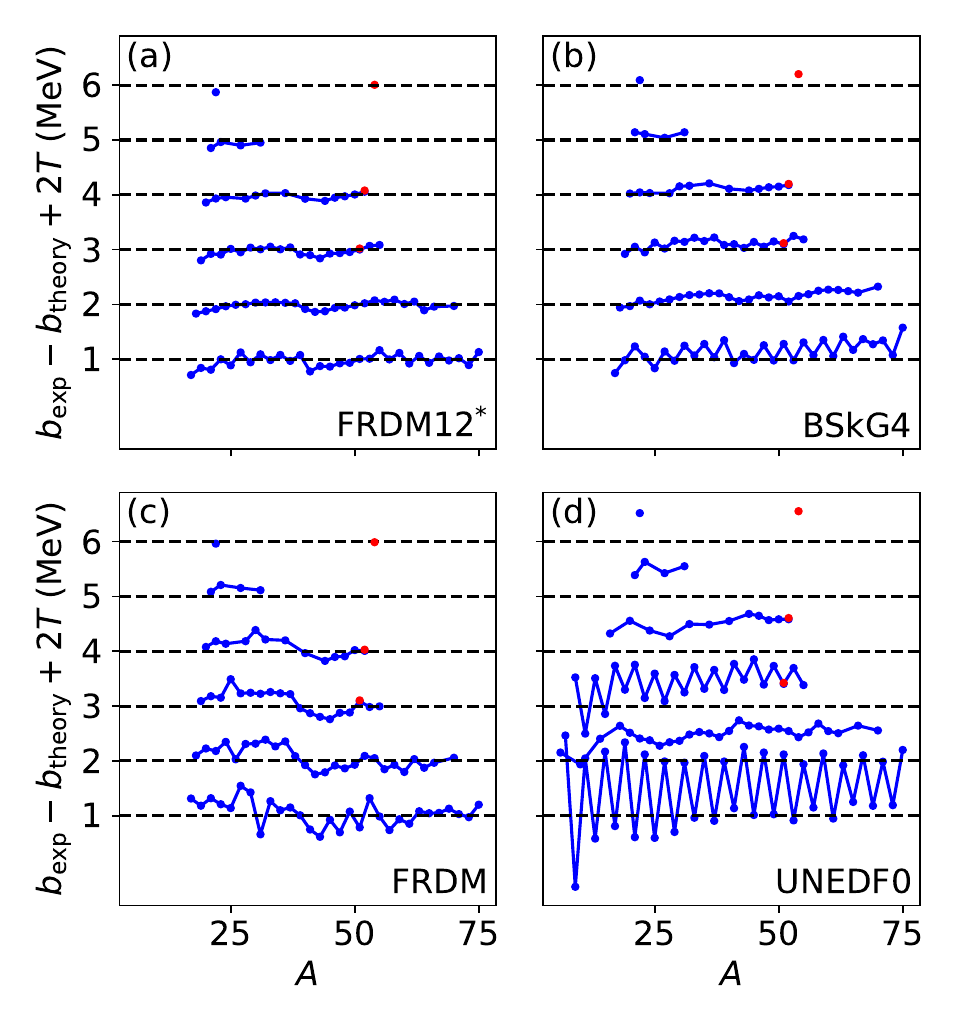}
\caption{Deviations between experimental and theoretical $b$-coefficients shifted by $2T$ as a function of mass number for (a) FRDM12*, (b) BSkG4, (c) FRDM and (d) UNEDF0. Data points containing experimental data from NUBASE2020 including recent mass measurements as reported in Refs.~\cite{PhysRevLett.133.222501, 14s5-17gj, PhysRevC.110.L031301, ffwt-n7yc} are depicted in blue while data points containing experimental data from this work are shown in red.}
\label{fig:bexp-btheory}
\end{figure}

This behavior is valid across the nuclear chart as illustrated in Figs.~\ref{fig:bvsA_fullchart} and~\ref{fig:bexp-btheory}. FRDM12 and BSkG4 generally reproduce the experimental data, shown in green, fairly well, whereas UNEDF0 predicts significantly smaller $b$ coefficients than observed experimentally. In addition, the theoretical results for UNEDF0 exhibit a pronounced odd--even staggering for the $T=1/2$, $3/2$, and $5/2$ nuclei, a feature that is not present in the experimental data. While the overall agreement of FRDM12 with experiment is good, a closer examination reveals that part of the remaining discrepancies originate from the treatment of the ground-state microscopic energies, $E_{\mathrm{mic}}$. FRDM12 employs different $E_{\mathrm{mic}}$ of the mirror nuclei (see Tab. 1 of Ref.~\cite{FRDM2012}), implying that the two members of a mirror pair have different deformations. When assuming the same $E_{\mathrm{mic}}$ for both mirror nuclei, so when neglecting the $E_{\mathrm{mic}}$ term for the FRDM12 calculations, the agreement with experimental data improves significantly. The corresponding results, labeled FRDM12*, are shown in Fig.~\ref{fig:FRDMstar} and the differences between experimental and theoretical $b$ coefficients in Fig.~\ref{fig:bexp-btheory}(a). The largest deviations between the experimental data and FRDM12* occur due to shell effects around $A=16$ and $A=40$, where the valence orbitals change from $0p$ to $0d1s$, and from $0d1s$ to $0f1p$, respectively, resulting in larger rms radii and lower Coulomb energies. For the $T=1/2$ nuclei, one can see some small odd-even oscillations for $A=16-40$, which are reproduced in $0d1s$ shell-model calculations that contain Coulomb pairing~\cite{PhysRevC.101.064312}.

\begin{figure}[t]
\includegraphics[width=\columnwidth]{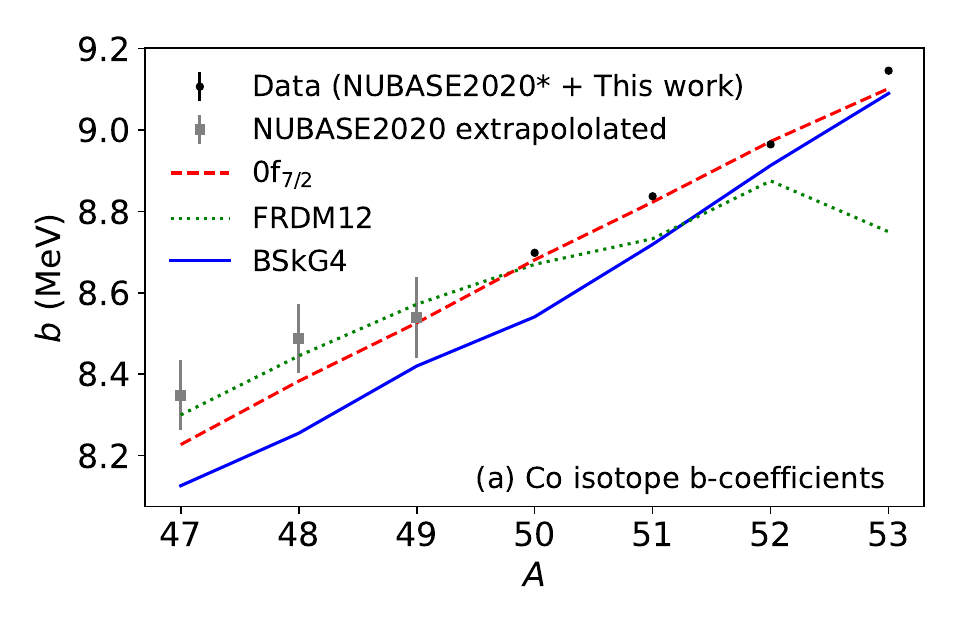}
\includegraphics[width=\columnwidth]{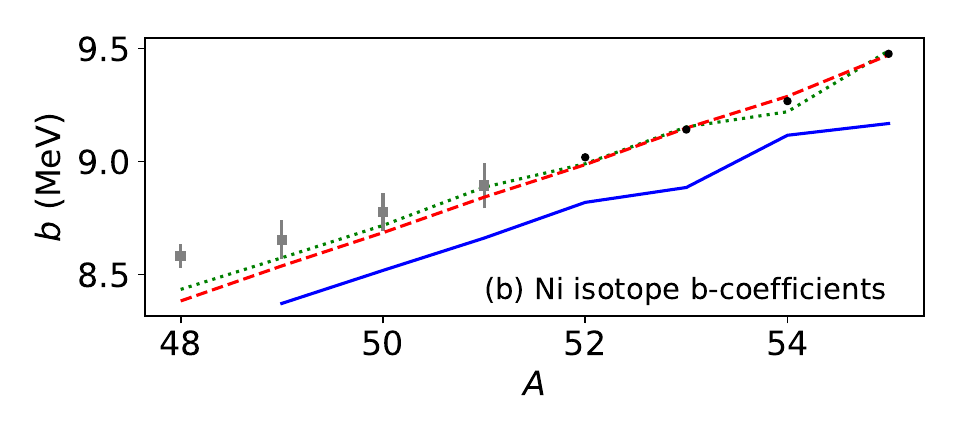}
\includegraphics[width=\columnwidth]{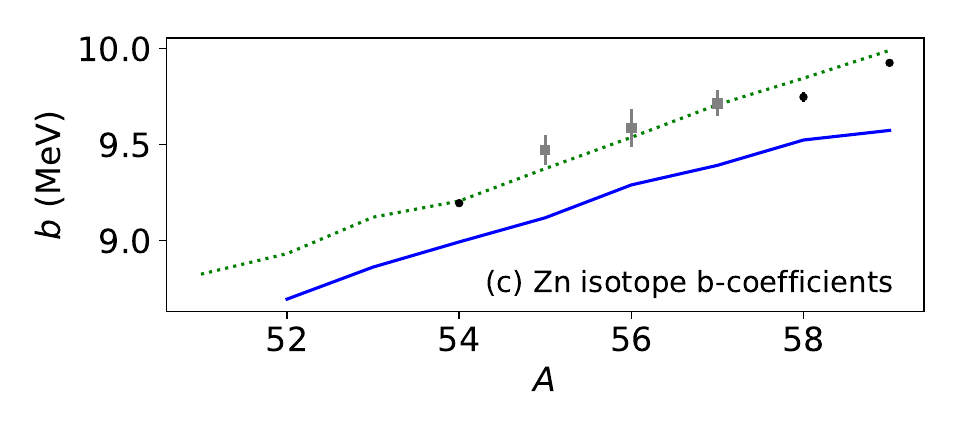}
\caption{The $b$-coefficients as a function of mass number for (a) cobalt, (b) nickel and (c) zinc isotopes. The black data points are either calculated based on the experimental data reported in NUBASE2020~\cite{Kondev_2021} or in this work with the exception of the mass excess values for $^{53,54}$Ni, which are taken from Ref.~\cite{Zhang2023}. The gray data points are the extrapolated values in NUBASE2020~\cite{Kondev_2021}. The red-dashed line corresponds to a local model based on a fit to displacement energies for nuclei in the $0f_{7/2}$ model space developed in 1979~\cite{BROWN197961}, the green-dotted line shows the FRDM12 model~\cite{moller2016nuclear} and the blue full line the BSkG4 model~\cite{BSKG4}.}
\label{fig:isotopes_bvsA}
\end{figure}

Another comparison is depicted in Fig.~\ref{fig:isotopes_bvsA}, which shows the $b$ coefficients of various theoretical models alongside the experimental data for the Co, Ni and Zn isotopic chains. While the BSkG4 model reproduces the general trend well, it yields substantially lower $b$ values. The FRDM12 model describes the data fairly well, except for noticeable deviation in the Co isotopic chain for $A>51$. We also included the regional $0f_{7/2}$ shell model predictions~\cite{BROWN197961} for Co and Ni isotopes. Developed in 1979, this model is based on a fit to displacement energies for nuclei in the $0f_{7/2}$ model space and was previously used to predict the Q value for di-proton decay of $^{48}$Ni~\cite{PhysRevC.43.R1513}. Among the considered models, it provides the best description of the experimental data. 

The evaluation of the $  d  $-coefficent of the IMME requires to know the binding energies of at least 4 members of the isobaric multiplet, such that $d$ can be obtained from a least-square minimization fit to the cubic form of the IMME.  For the $A=51$, $T=3/2$ and $A=52$, $T=2$ and $A=54$, $T=3$ multiplets, the mass excess values of the IAS of $^{51}$Co, $^{52}$Ni and $^{54}$Zn correspond directly to the measured ground state masses in this work. For the $A=51$, $T=5/2$ multiplet, the IAS excitation energy of $^{51}$Co was determined to be $E_\mathrm{IAS} = 6702(18)$~keV using the procedure described in~\cite{DOSSAT200718} and combining the values reported therein with our measured ground-state mass of $^{51}$Co. For all the other IAS, we adopted the values listed in NUBASE2020, with the exception of the ground-state masses of $^{27}$P and $^{22}$Al, which were taken from ~\cite{PhysRevC.108.065802, PhysRevLett.132.152501} and the $E_\mathrm{IAS}$ of $^{27}$Si, for which we used 6638(1)~keV~\cite{PhysRevC.94.065806}. Table~\ref{tab: IASTable} lists all available mass excess values for the ground state and IAS as well as the excitation energy $E_\mathrm{IAS}$ for all members of the $A=51$, $52$ and $54$ multiplets. 

\begin{table*}[t] 
\centering
\caption{Available mass excess values for the ground state $\mathrm{ME}_{\mathrm{gs}}$, excitation energy $E_\mathrm{IAS}$, mass excess value for the IAS $\mathrm{ME}_{\mathrm{IAS}}$, and binding energy for the IAS  $\mathrm{BE}_{\mathrm{IAS}}$ for the respective members of the $A=51$, $52$ and $54$ multiplets. The data is taken from this work, Ref.~\cite{DOSSAT200718} and NUBASE2020~\cite{Kondev_2021}. Not available experimental data is indicated with N/A. }
\vspace{\baselineskip}
\renewcommand{\arraystretch}{1.25}
\setlength{\tabcolsep}{12pt}
\begin{tabular}{c l c S S S S}
\hline
&&
\multicolumn{1}{c}{$T_z$} &
\multicolumn{1}{c}{$\mathrm{ME}_{\mathrm{gs}}$ (keV)} &
\multicolumn{1}{c}{$E_{\mathrm{IAS}}$ (keV)} &
\multicolumn{1}{c}{$\mathrm{ME}_{\mathrm{IAS}}$ (keV)} &
\multicolumn{1}{c}{$\mathrm{BE}_{\mathrm{IAS}}$ (keV)} \\\\
\hline
$A=51$, $T=3/2$ & $^{51}$Co &-3/2 & -27375.1(5.7) & \multicolumn{1}{c}{--} & -27375.1(5.7) & 417888.9(5.7)  \\ 
&$^{51}$Fe &-1/2 & -40189.2(1.4) & \multicolumn{1}{c}{N/A}  & \multicolumn{1}{c}{N/A}   & \multicolumn{1}{c}{N/A}  \\ 
&$^{51}$Mn & 1/2  & -48243.2(3) & 4450.6(1.5)  & -43792.6(1.5)  & 435871.1(1.5) \\ 
&$^{51}$Cr & 3/2 & -51450.71(17) & \multicolumn{1}{c}{--} & -51450.71(17)  & 444311.60(17)\\ 
\hline
$A=51$, $T=5/2$ & $^{51}$Ni &-5/2 & \multicolumn{1}{c}{N/A}  & \multicolumn{1}{c}{--}  & \multicolumn{1}{c}{N/A}  & \multicolumn{1}{c}{N/A}   \\ 
&$^{51}$Co &-3/2 & -27375.1(5.7) & 6702(18) & -20673(19) & 411187(19)  \\ 
&$^{51}$Fe &-1/2 & -40189.2(1.4) & \multicolumn{1}{c}{N/A}  & \multicolumn{1}{c}{N/A}   & \multicolumn{1}{c}{N/A}  \\ 
&$^{51}$Mn & 1/2  & -48243.2(3) & \multicolumn{1}{c}{N/A}  & \multicolumn{1}{c}{N/A}  & \multicolumn{1}{c}{N/A}   \\ 
&$^{51}$Cr & 3/2 & -51450.71(17) & 6613(5) & -44837.7(5.0)  & 437698.6(5.0) \\ 
&$^{51}$V & 5/2 & -52203.11(10) & \multicolumn{1}{c}{--}  & -52203.11(10) & 445846.35(10) \\ 
\hline
$A=52$, $T=2$ & $^{52}$Ni &-2 & -22474.8(2.2) & \multicolumn{1}{c}{--}  & -22474.8(2.2) & 420277.6(2.2) \\ 
&$^{52}$Co &-1 & -34344(5) & 2924(9) & -31420(8)  & 430005(8) \\ 
&$^{52}$Fe &0 & -48332.10(18) & 8556(6) & -39776(6) &439144(6) \\ 
&$^{52}$Mn & 1  & -50711.39(13) & 2926(5) & -47785(5)  & 447934.9(5.0) \\ 
&$^{52}$Cr & 2 & -55419.51(11) & \multicolumn{1}{c}{--}  & -55419.51(11) & 456351.72(11)  \\ 
\hline
$A=54$, $T=3$ & $^{54}$Zn &-3 & -6463(42) & \multicolumn{1}{c}{--}  & -6463(42) & 418844(42) \\ 
&$^{54}$Cu &-2 & \multicolumn{1}{c}{N/A}  & \multicolumn{1}{c}{N/A}  & \multicolumn{1}{c}{N/A}   & \multicolumn{1}{c}{N/A}  \\ 
&$^{54}$Ni & -1 & -39278(5) & \multicolumn{1}{c}{N/A} & \multicolumn{1}{c}{N/A} & \multicolumn{1}{c}{N/A}  \\ 
&$^{54}$Co & 0  & -48010.1(4) & \multicolumn{1}{c}{N/A}  & \multicolumn{1}{c}{N/A}   & \multicolumn{1}{c}{N/A}    \\ 
&$^{54}$Fe & 1 & -56254.6(3) & 14868(20) & -41387(20)  &456897(20) \\ 
&$^{54}$Mn & 2  & -55558.2(1.0) & 6146.4(3.0) & -49411.9(2.8) & 465704.4(2.8) \\ 
&$^{54}$Cr & 3 & -56935.38(13) & \multicolumn{1}{c}{--}  & -56935.38(13) & 474010.23(13) \\ \hline 
\end{tabular}
\label{tab: IASTable}
\end{table*}

\begin{figure}[t]
\includegraphics[width=\columnwidth]{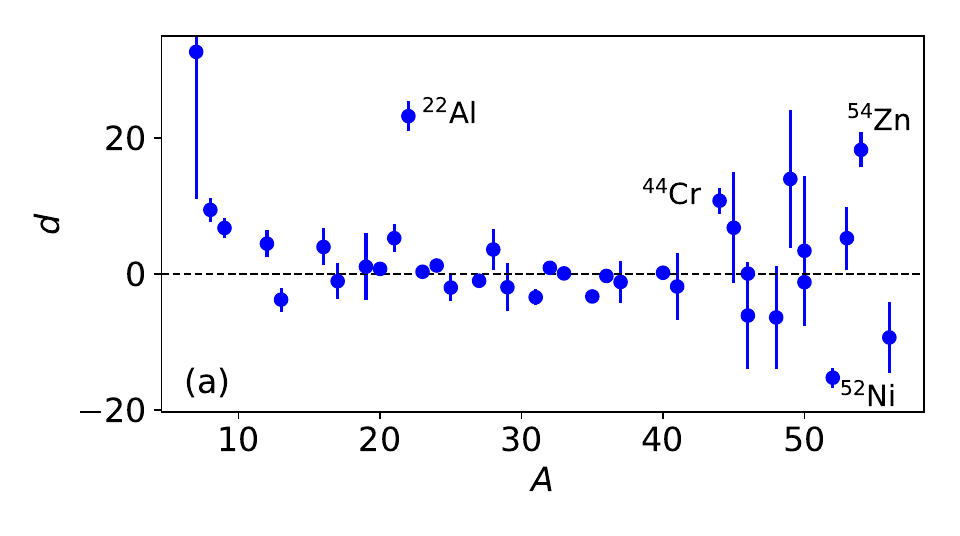}
\includegraphics[width=\columnwidth]{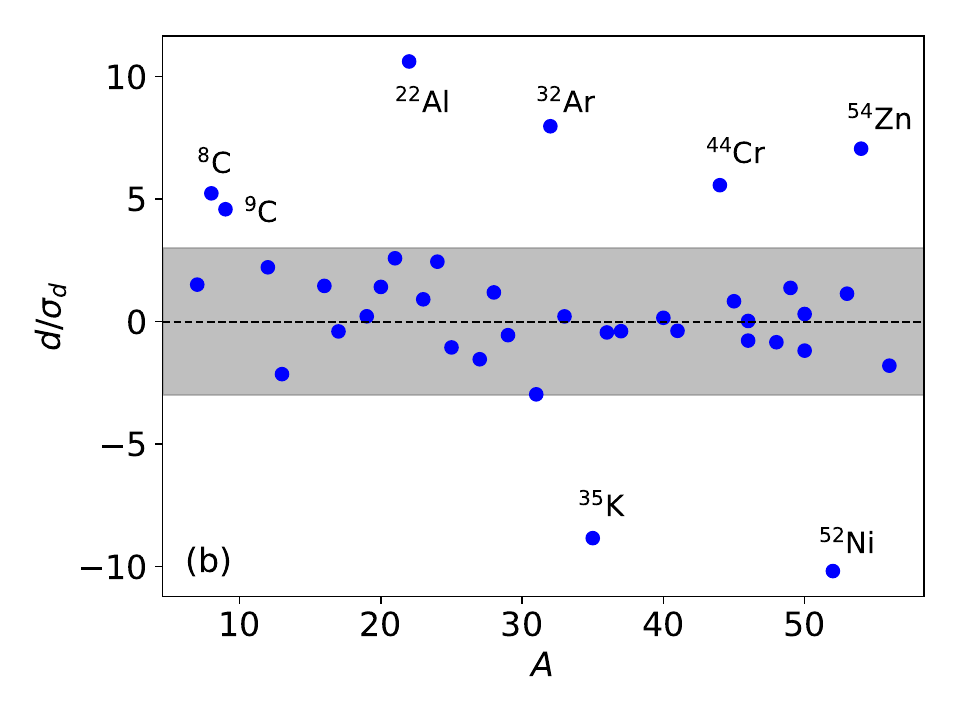}
\caption{(a) The $d$-coefficients as a function of mass number $A$. (b) The $d$-coefficients divided by their error $\sigma_d$ as a function of $A$. The experimental data used to determine the $d$ coefficients is taken from this work, NUBASE2020 and Refs.~\cite{DOSSAT200718, PhysRevC.108.065802, PhysRevLett.132.152501, PhysRevC.94.065806}. All data points within the gray shaded band in panel (b) deviate by less than $\pm 3\sigma$ from 0. The text labels indicate the proton-richest members of some of the multiplets.}
\label{fig:d coeff}
\end{figure}

All obtained $d$ coefficients and their deviation from 0 are shown in Fig.~\ref{fig:d coeff} as function of the mass number $A$. Whereas the deviation from 0 is typically within $\pm 3\sigma$, it is about $5\sigma$ for $A=8$ and 9, $7\sigma$ for $A=54$ and between $8\sigma$ and $11\sigma$ for $A=22, 32,35,44$ and 52. The large deviation for $A=32$ might be due to an erroneous ground state mass of $^{32}$Cl, a wrong excitation energy of one of the IAS or extremely strong isospin mixing~\cite{PhysRevC.80.051302}. For $A=35$, the data for the excited $^{35}$Cl levels or the ground state mass of $^{35}$Ar is put into question~\cite{PhysRevC.76.024308}. For $A=44$, the IAS in $^{44}$V might have been misidentified~\cite{PhysRevC.102.054311}. 

In Ref.~\cite{PhysRevC.102.054311} the $d$ coefficient for the $  A=52  $, $  T=2  $ quintet was in agreement with 0. However, with our substantially more precise mass measurement of $^{52}$Ni the $d$ coefficient is $10.3 \sigma$ off. This represents the second largest deviation of the experimental $d$ coefficients from 0 amongst all multiplets, see Fig.~\ref{fig:d coeff}. Employing the cubic form of the IMME, we obtain a $d$ coefficient of $-15.4(1.5)$~keV. For the quartic form, the $d$ coefficient is $-17.9(1.6)$~keV and the $e$ coefficient is $11.2(2.2)$~keV. 
In addition to the fit of the IMME, $  d  $ can also be evaluated based on 
\begin{equation}
     b  = \frac{ \mathrm{BE}(^{52} \mathrm{Cr}) - \ \mathrm{BE}(^{52}\mathrm{Ni})    }{4}, 
\end{equation}
\begin{equation}
     b'  = \frac{\mathrm{BE}(^{52}\mathrm{Mn, IAS}) - \mathrm{BE}(^{52}\mathrm{Co, IAS})}{2},
\end{equation}
and
\begin{equation}
     d = \frac{b'-b}{3} .
\end{equation}

 For the experimental data we obtain $  b = 9.0185(6) $, $  b'  = 8.9648(47)$~MeV, and $  d = -17.9(1.6)  $~keV, in agreement with the value obtained from the quartic form of the IMME. Conversely, for the $  f_{7/2}  $ model $  b = b'   = 8.986$~MeV
and $  d = 0  $, indicating no signature of a breakdown of the IMME. 

The large deviation of the experimental $d$ coefficient from 0 might originate from a misidentification of one of the IAS or a wrongly reported mass or excitation energy $E_\mathrm{IAS}$. Mass measurements of $^{52}$Co, performed at IGISOL~\cite{Nesterenko_2017} and CSRe~\cite{PhysRevLett.117.182503}, deviate by $2.8\sigma$. The $d$ coefficient in the cubic form of the IMME is $-14.0(1.8)$~keV when taking the CSRe mass value and $-11.8(1.8)$~keV when taking the IGISOL value. Only when neglecting the IAS of $^{52}$Co entirely, the $d$ coefficient is $-6.7(2.3)$~keV and hence within $\pm 3\sigma$ from 0. When neglecting the IAS of $^{52}$Fe, $^{52}$Mn or $^{52}$Cr, the $d$ value is still larger than $8 \sigma$. A misassignment of the IAS of $^{52}$Co may therefore be the cause of the breakdown of the IMME for $A=52$. It should be noted that the identification of IAS in measurements of $\beta p$ decays with low branching ratios has recently been put into question~\cite{SU2016323}. 

For the $  A=54  $, $  T=3  $ septet, the IAS mass values of 4 members are known. The $d$ coefficient can be obtained from the cubic fit of the IMME with 18.2(2.6)~keV, leading to a $7 \sigma$ deviation of $d$ from 0. Excluding the mass of the ground state of $^{54}$Zn and obtaining $a,b$ and $c$ from the fit to the quadratic form of the IMME, the IMME prediction of the mass excess of $^{54}$Zn would be -4273(309)~keV. This value deviates by 2190(312)~keV from the indirect mass determination of -6463(42)~keV based on the ground state mass of $^{52}$Ni and the measured two-proton decay energy of $^{54}$Zn, see the Results Section. It also deviates by 1427(379)~keV from the extrapolated mass in AME2020 given by -5700(220)~keV. Here again, a misidentification of one of the IAS, presumably $^{54}$Fe, provides the most plausible explanation for the breakdown of the IMME as suggested in~\cite{Li_2022}.

For the $A=51, T=3/2$ quartet and for the $A=51, T=5/2$ sextet, the IAS mass values of only three members are known, such that the $d$ coefficient cannot be obtained. Using the quadratic form of the IMME, we extract the corresponding $a,b$ and $c$ coefficients for the $A=51, T=5/2$ sextet from a least-square fit and use them to predict the ground-state mass of $^{51}$Ni. The resulting value of -11928(33)~keV is in good agreement with the extrapolated value of -11650(500)~keV reported in the AME2020. A Penning trap mass measurement of $^{51}$Ni would obtain the $d$ coefficient for $A=51$, which will be within reach at FRIB in the near future.

One must consider the reliability of the experimental data that was used to determine the energies of the IAS, particularly the assignment of IAS, as well as their possible mixing with states with lower isospin. Theoretical models for a non-zero value of $  d  $ should be considered in future work.

\section{V. Conclusions}
We performed Penning-trap mass measurements of $^{51}$Co and $^{52}$Ni, improving the mass precision by factors of 2 and 37. Using the mass of $^{52}$Ni together with the known two-proton decay energy of $^{54}$Zn, we also determined the mass of $^{54}$Zn. 

Our new mass values enable stringent tests of the IMME and of isospin-symmetry-breaking effects in the proton-rich $fp$ shell. In particular, comparisons of experimental $b$ coefficients with global nuclear models demonstrate that models omitting the Coulomb-exchange term generally reproduce the experimental data better than conventional Skyrme-type EDF approaches employing the Slater approximation. The regional $0f_{7/2}$ model provides the best overall agreement for the Co and Ni isotopic chains considered in this work.

Our improved mass precision for $^{52}$Ni results in a significant non-zero cubic coefficient for the $A=52$, $T=2$ multiplet. Also the $A=54$, $T=3$ multiplet exhibits a substantial deviation from the quadratic form of the IMME. These deviations belong to the largest observed breakdowns of the IMME and strongly suggest either unresolved experimental issues, such as IAS misassignments or inaccurate binding energies of the IAS, or the need for additional theoretical contributions beyond the standard quadratic form.
Future high-precision mass measurements in the $fp$ shell and a reevaluation of IAS will be essential to clarify the origin of the observed IMME breakdowns. 

\section{Acknowledgements}
This material is based upon work supported by the U.S. Department of Energy, Office of Science, Office of Nuclear Physics and used resources of the Facility for Rare Isotope Beams (FRIB) Operations, which is a DOE Office of Science User Facility under Award Number DE-SC0023633.
This work was conducted with the support of Michigan State University, the US National Science Foundation under contract no. PHY-2111185 and PHY-2110365, the DOE, Office of Nuclear Physics under contract no. DE-AC02-06CH11357, DE-AC02-05CH11231, DE-SC0022538, and DE-SC0023688.   
This material is based upon work supported by the Department of Energy, National Nuclear Security Administration under Award Number DE-NA0004252.
S.E.C. acknowledges support from the DOE NNSA SSGF under DE-NA0003960.


%\bibliography{PRC.bib} 
\vspace{1mm}

\end{document}